\documentclass[conference]{IEEEtran}
\IEEEoverridecommandlockouts

\makeatletter
\def\markboth#1#2{\def\leftmark{\@IEEEcompsoconly{\sffamily}\MakeUppercase{\protect#1}}%
\def\rightmark{\@IEEEcompsoconly{\sffamily}\MakeUppercase{\protect#2}}}
\makeatother

\usepackage{graphicx}
\usepackage{epsfig}
\usepackage{epstopdf}
\usepackage{amstext}
\usepackage{amssymb}
\usepackage{amsmath}
\usepackage{amsthm}
\usepackage[font=footnotesize]{caption}
\usepackage{subcaption}
\usepackage{cite}
\usepackage{float}
\usepackage{array}
\usepackage{url}
\usepackage{flexisym}
\usepackage{soul}
\usepackage{color}
\usepackage{bbm}
\usepackage{dsfont}
\usepackage[T1]{fontenc}
\usepackage[utf8]{inputenc}
\usepackage{authblk}
\usepackage{gensymb}
\usepackage[ruled,vlined,linesnumbered]{algorithm2e}

\newcommand{\reminder}[1]{} % use this in the final version

\IEEEoverridecommandlockouts
\DeclareMathOperator*{\argmax}{arg\,max}
\DeclareMathOperator*{\argmin}{arg\,min}

\begin{document}

\title{Scheduling and Power Allocation in Self-Backhauled Full Duplex Small Cells\vspace{-4mm}}
\author{Sanjay Goyal\thanks {This work is funded by NSF (CNS-1527750) and by the New York State Center for Advanced Technology in Telecommunications (CATT).}}
\author{Pei Liu}
\author{Shivendra Panwar\vspace{-4mm}}
\affil{NYU Tandon School of Engineering, Brooklyn, NY, USA \vspace{-4mm}}
\affil{{\{sanjay.goyal, peiliu, panwar\}@nyu.edu}\vspace{-6mm}}
\maketitle
\begin{abstract}
Full duplex (FD) communications, which increases spectral efficiency through simultaneous transmission and reception on the same frequency band, is a promising technology to meet the demand of next generation wireless networks. In this paper, we consider the application of such FD communication to self-backhauled small cells. We consider a FD capable small cell base station (BS) being wirelessly backhauled by a FD capable macro-cell BS. FD communication enables simultaneous backhaul and access transmissions at small cell BSs, which reduces the need to orthogonalize allocated spectrum between access and backhaul. However, in such simultaneous operations, all the links experience higher interference, which significantly suppresses the gains of FD operations. We propose an interference-aware scheduling method to maximize the FD gain across multiple UEs in both uplink and downlink directions, while maintaining a level of fairness between all UEs. It jointly schedules the appropriate links and traffic based on the back-pressure algorithm, and allocates appropriate transmission powers to the scheduled links using Geometric Programming. Our simulation results show that the proposed scheduler nearly doubles the throughput of small cells compared to traditional half-duplex self-backhauling. 
\end{abstract}


%%%%%%%%%%%%%%%%%%%%%%%%%
%%%%%%%%%%%%%%%%%%%%%%%%%

\section{Introduction}\label{sec1}

The demand for wireless data has been increasing at a rapid pace. The next generation mobile networks that will be deployed at 2020s aim at an up to thousand times increase in traffic as compared to traffic in 2010~\cite{Metis,NGMN_5G}. Such traffic demand poses several challenges to the network, including the air interface. Currently, the solutions being pursued fall into two categories: more available spectrum and higher spectrum efficiency. The latter includes new radio link design such as waveforms and channel coding enhancements; network architecture evolution towards small cells, heterogeneous networks, and multi-cell cooperation. For 5G, dense small cell deployments will be a key feature  to enable efficient spectral reuse. However, increasing the number of small cells will impose a much higher demand on the backhaul network. Wireless backhaul is essential to provide connectivity to small cells, since it is much more cost-efficient compared to fiber based backhaul for the last hundred meters ~\cite{ekram_self_backhaul_magazine, alto_self_backhaul}. 

Recent advances in antenna and RF circuit design have greatly reduced the crosstalk between the transmitter and receiver circuits on a wireless device, which enables radios to transmit and receive on the same frequency at the same time (\emph{Full Duplex (FD) Radio})~\cite{survey_JSAC, survey_kim}. Such FD radio can in addition provide more efficient spectrum reuse. Using FD radio for simultaneous backhaul and access transmissions at small cell BSs, i.e., self-backhauling, reduces the need to orthogonalize allocated spectrum between access and backhaul. Small cell BS can schedule its downlink and uplink traffic simultaneously using the same channel. 

In this paper, we consider a FD capable small cell BS being wirelessly backhauled by a FD capable macro-cell BS. Using FD operations simultaneously for both backhaul and access links implies that uplink and downlink access or backhaul links experience higher interference compared to the half-duplex (HD) operations using orthogonal resources. The high interference raises several questions regarding the potential performance of FD operation. The actual gain from FD operation will strongly depend on link geometries, propagation effects in mobile channels, and power levels at the nodes involved in transmissions. 

In this paper, we focus on the design of the combination of an interference-aware scheduler and power control algorithm that maximizes the FD gain across multiple UEs in both uplink and downlink directions, while maintaining a level of fairness between all UEs. In such a system, FD gain can be achieved by simultaneous transmission in backhaul and access links, where the the extra FD interference would be treated as noise. The scheduler is a hybrid scheduler in the sense that it will exploit FD transmissions only when it is advantageous to do so. Otherwise, when the interference is too strong, or traffic demands dictate it, it might conduct HD operations. 

Exploiting FD radio for providing backhauling to small cells have also been investigated recently in~\cite{Ganti_self-backhaul_arxiv, erkam_self_backhaul_journal_16,korpi_self_backhaul_16,rahmati_price_based_self_back,liang_self_backhaul}. In \cite{Ganti_self-backhaul_arxiv}, Sharma \emph{et al.}~showed the downlink coverage and throughput trade-off of the FD self-backhauled small cell using stochastic geometry. They showed that the downlink rate in such networks could be close to double that of a conventional TDD/FDD self-backhauling network, but at the expense of reduced coverage due to higher interference under FD operations. Similarly, Tabassum \emph{et al.} in~\cite{erkam_self_backhaul_journal_16} derived the downlink coverage probability of the FD self-backhauled small cell and showed the impact of additional interference due to FD operations. They also discussed the need for interference management solutions such as employing hybrid HD and FD operation and power control, which we also consider in our joint scheduling and power allocation policy in this paper. Korpi \emph{et al.}~\cite{korpi_self_backhaul_16} derived achievable sum-rates for both downlink and uplink, assuming large arrays of antennas at the FD self-backhauled small cell BS, to facilitate efficient beamforming and self-interference nulling at its own receiver. They also allowed device to device transmission and consider the case where small cell BS relays the traffic inside the cell without forwarding it to the backhaul link. They showed that the highest sum-rate is usually achieved when the small cell BS acts as a FD relay between the user equipment (UEs) and the backhaul node. In~\cite{liang_self_backhaul}, Li \emph{et al.} considered the case of massive MIMO at the macro BS serving several small cell BS with the support of self-backhauling and derived the downlink and uplink throughput using zero-forcing beamforming and decoding, respectively.  

In~\cite{rahmati_price_based_self_back}, a game theory based approach using the Stackelberg game model is used to allocate the powers to UEs in the downlink direction, considering the simultaneous backhaul and access dowlink transmission at the small cell BS. The limitation of the existing works discussed above is that they either do not consider multi-UE diversity gain, which comes through scheduling of the appropriate UEs with power adjustments to mitigate interference, or they do not consider the use of FD capability in all possible combinations including small cell simultaneous uplink and downlink access transmissions, both of which are considered in this paper. The key contributions of this paper are:

\begin{itemize}
\item
A capacity analysis is presented to compare the performance of the system with FD and HD operations under different propagation conditions. 
\item
A joint uplink and downlink scheduler is considered which schedules the appropriate traffic in each direction so that all the queues in the system remain stable.
\item
The scheduler jointly optimizes UE selection and power allocation such that the maximum throughput gain is achieved while maintaining a level of fairness among the UEs.
\item
The scheduler uses the FD capability of the small cell BS for all possible FD transmissions, i.e., simultaneous backhaul and access uplink/downlink transmissions, and simultaneous uplink and downlink access/backhaul transmissions.
\end{itemize}

The remaining part of the paper is organized as follows: Section~\ref{sec:SB_SM} describes the system model and problem formulation.
A capacity analysis to compare the performance of FD and HD operations is presented in Section~\ref{sec:SB_Comp}. The joint scheduling
and power allocation method is given in Section~\ref{sec:SB_sched}. Section~\ref{sec:SB_PE} contains simulation details and performance
results for the proposed FD scheduling algorithms. Conclusions are discussed in Section~\ref{sec:SB_conc}.

\section{System Model}\label{sec:SB_SM}
We consider a single macrocell ($\text{M}$), which provides wireless backhaul service to a single small cell ($\text{S}$) deployed in its coverage area. There are $N (\geq 1)$ UEs associated with the small cell. Each of the UEs communicate with the small cell BS for both downloading and uploading its data. Furthermore, we assume traffic to/from the UEs associated with the macrocell, and traffic to/from the UEs associated with the small cell use orthogonal channel resources. Thus there would be no interference between them, and this paper will focus only on the scheduling and power adaptation for traffic to and from the small cell UEs. We assume that both the macro-cell BS and small cell BS maintain a separate  pair of queues for each UE for uplink and downlink traffic. The arrival traffic is first buffered at both the BSs and transmitted per the scheduling decisions. Both the macro-cell BS and the small cell BS are FD capable. Due to the significant cost and power requirements of FD circuits, we envision none of the UEs can transmit and receive at the same time. However, as we will show in our simulation, such a limitation will not significantly reduce full duplex gains.

\begin{figure}
\centering
\includegraphics[width = 2.4 in] {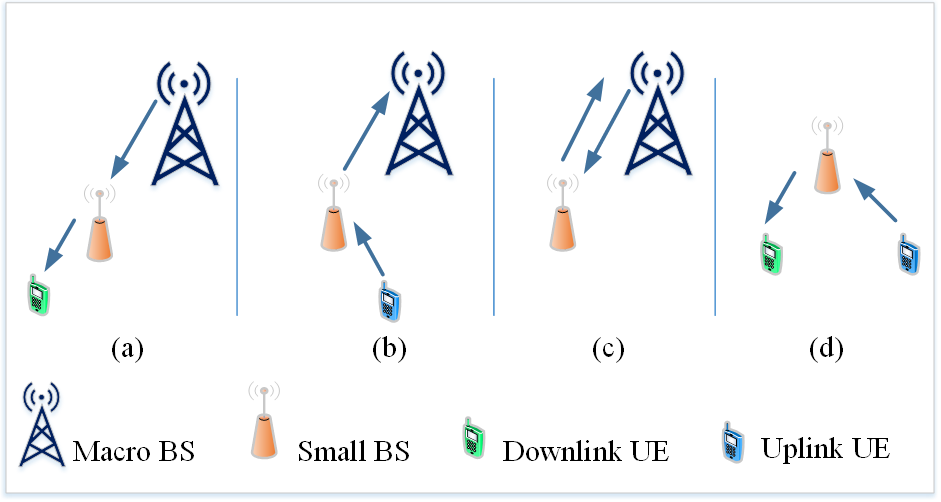}
\vspace{-2mm}
\caption{Full Duplex Transmission Modes.}
\label{fig:SB_diff_FDM}
\vspace{-4mm}
\end{figure}

\begin{figure}
\centering
\includegraphics[width = 2.4 in] {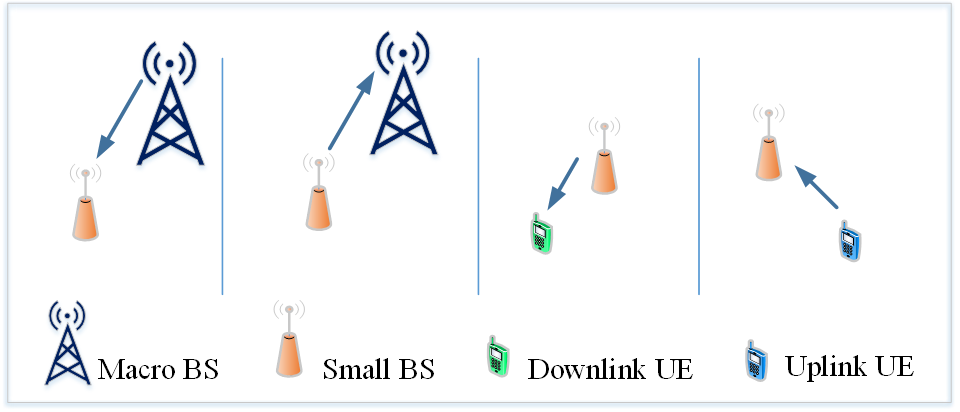}
\vspace{-2mm}
\caption{Half Duplex Transmission Modes.}
\label{fig:SB_diff_HDM}
\vspace{-7mm}
\end{figure}

Based on the above assumptions, Fig.~\ref{fig:SB_diff_FDM} illustrates all the possible full duplex transmission modes. Fig.~\ref{fig:SB_diff_FDM}a shows \textit{FD Downlink (FDD)} mode, where the simultaneous transmission from the macro to small cell, and the small cell to a downlink UE occur on the same channel. Fig.~\ref{fig:SB_diff_FDM}b shows \textit{FD Uplink (FDU)} mode, where the simultaneous transmission from an uplink UE to small cell, and the small cell to macro-cell can be scheduled on the same channel. In \textit{FD Backhaul (FDB)} mode, simultaneous uplink and downlink transmission occur between the small and macro-cell as shown in Fig.~\ref{fig:SB_diff_FDM}c. Similarly, in \textit{FD Access (FDA)} mode, as shown in Fig.~\ref{fig:SB_diff_FDM}d, the small cell schedules an uplink UE and a downlink UE for simultaneous uplink and downlink access. In a legacy HD system, each channel can only be used for a single transmission, and Fig.~\ref{fig:SB_diff_HDM} illustrates all the possible modes.

The purpose of our scheduler is to identify one of the transmission modes above that maximizes the spectrum efficiency based on dynamic channel and traffic conditions. The scheduler does not have to operate the system in full duplex mode ( Fig.~\ref{fig:SB_diff_FDM}) in every time slot; when it is more efficient to operate in HD mode, it will employ one of the modes in Fig.~\ref{fig:SB_diff_HDM}. The detailed scheduling algorithm can be found in Section~\ref{sec:SB_sched}.

Assume that at time slot $t$, in FDD mode, the macro-cell BS transmits the signal $x(t)$ to the small cell BS and the small cell BS transmits $y(t)$ to its selected downlink UE  in time slot $t$, i.e., $\text{D} \in \{1,2,\cdots,N\}$. Hence, the received signals at the small cell BS and at the selected downlink UE are given, respectively, by
\vspace{-4mm}
\begin{equation}\label{eq:SB_rx_sig_sBS_FDD}
s(t) = h_{\text{MS}} x(t) + h_{\text{SS}} y(t) + n_{\text{S}},
\end{equation}
\begin{equation}\label{eq:SB_rx_sig_sD_FDD}
d(t) = h_{\text{SD}} y(t) + h_{\text{MD}} x(t) + n_{\text{D}},
\end{equation}
where $h_{\text{MS}}$, $h_{\text{SD}}$, and $h_{\text{MD}}$ are used to denote the complex channel response between the macro-cell BS and the small cell BS, between the small cell BS and the downlink UE $\text{D}$, and between the macro-cell BS and the downlink UE $\text{D}$, respectively. It includes path loss, small-scale fading and shadowing. The self-interference channel at the small cell BS is denoted by $h_{\text{SS}}$, which includes the cancellation. We model the transmitted symbols ($x(t)$, $y(t)$) as independent random variables with zero mean and variance, $\mathbb{E}\{|x(t)|^2\} \overset{\Delta}{=} p_{\text{M}}(t) \geq 0$, and $\mathbb{E}\{|y(t)|^2\} \overset{\Delta}{=} p_{\text{S}}(t) \geq 0$. The notation $n_{\text{S}}$ and $n_{\text{D}}$ denote the additive noise at the small cell BS and the downlink UE $\text{D}$, treated as complex Gaussian random variable with variances $\mathcal{N}_{S}/2$ and $\mathcal{N}_{D}/2$, respectively. 

Thus, in FDD mode, the signal to interference plus noise (SINR) for the small cell BS and the downlink UE $\text{D}$ are given by, respectively,
\begin{equation}\label{SB_sinr_BS_D_FDD}
\text{SINR}_{\text{S}}^{\text{FDD}} = \frac{p_{\text{M}}(t)G_{\text{MS}}}{ p_{\text{S}}(t)\gamma_{\text{S}} + \mathcal{N}_{\text{S}}},~\text{SINR}_{\text{D}}^{\text{FDD}} = \frac{p_{\text{S}}(t)G_{\text{SD}}}{ p_{\text{M}}(t)G_{\text{MD}} + \mathcal{N}_{\text{D}}}.
\end{equation}

In the above equations, $G_{m,n} = |h_{m,n}|^2~\forall m,n$. The residual self-interference is modeled as Gaussian noise, the power of which equals the difference between the transmit power of the BS and the assumed amount of self-interference cancellation. $\gamma_{\text{S}}$ denotes the self-interference cancellation (SIC) level at the small cell BS. Similarly, SINRs can be defined for other modes. In FDU mode, SINR at the small cell BS and at the macro-cell BS are given by, respectively,
\begin{equation}\label{SB_sinr_BS_D_FDU}
\text{SINR}_{\text{S}}^{\text{FDU}} = \frac{p_{\text{U}}(t)G_{\text{US}}}{ p_{\text{S}}(t)\gamma_{\text{S}} + \mathcal{N}_{\text{S}}},~\text{SINR}_{\text{M}}^{\text{FDU}} = \frac{p_{\text{S}}(t)G_{\text{SM}}}{ p_{\text{U}}(t)G_{\text{UM}} + \mathcal{N}_{\text{M}}}.
\end{equation}

In FDB mode, SINR at the small cell BS and at the macro-cell BS are given by, respectively,
\begin{equation}\label{SB_sinr_BS_D_FDB}
\text{SINR}_{\text{S}}^{\text{FDB}} = \frac{p_{\text{M}}(t)G_{\text{MS}}}{ p_{\text{S}}(t)\gamma_{\text{S}} + \mathcal{N}_{\text{S}}},~\text{SINR}_{\text{M}}^{\text{FDB}} = \frac{p_{\text{S}}(t)G_{\text{SM}}}{ p_{\text{M}}(t)\gamma_{\text{M}} + \mathcal{N}_{\text{M}}}.
\end{equation}

In FDA mode, SINR at the downlink UE $\text{D}$ and at the small cell BS are given by, respectively,
\begin{equation}\label{SB_sinr_BS_D_FDA}
\text{SINR}_{\text{D}}^{\text{FDA}} = \frac{p_{\text{S}}(t)G_{\text{SD}}}{ p_{\text{U}}(t)G_{\text{UD}} + \mathcal{N}_{\text{D}}},~\text{SINR}_{\text{S}}^{\text{FDA}} = \frac{p_{\text{U}}(t)G_{\text{US}}}{ p_{\text{S}}(t)\gamma_{\text{S}} + \mathcal{N}_{\text{M}}}.
\end{equation}

In the above equations $\gamma_{\text{M}}$ and  $\mathcal{N}_{\text{M}}$ represent the SIC level and noise power, respectively, at the macro-cell BS; $\text{U}$ is used to represent the selected uplink UE, which means $p_{\text{U}}(t)$ represents the variance of the transmitted symbol from the uplink UE $\text{U}$. Further, in the case of HD transmissions, SNR expressions at the small cell BS in downlink (HDD) and uplink (HDU) will be given by, respectively,
\begin{equation}\label{SB_sinr_BS_HD_sBS}
\text{SNR}_{\text{S}}^{\text{HDD}} = \frac{p_{\text{M}}(t)G_{\text{MS}}}{ \mathcal{N}_{\text{D}}},~\text{SNR}_{\text{S}}^{\text{HDU}} = \frac{p_{\text{U}}(t)G_{\text{US}}}{ \mathcal{N}_{\text{S}}}.
\end{equation}

Similarly, SNRs at the downlink UE $\text{D}$ and the macro BS will be given by, respectively,
\begin{equation}\label{SB_sinr_BS_HD_UE_MBS}
\text{SNR}_{\text{D}}^{\text{HDD}} = \frac{p_{\text{S}}(t)G_{\text{SD}}}{ \mathcal{N}_{\text{D}}},~\text{SNR}_{\text{M}}^{\text{HDU}} = \frac{p_{\text{S}}(t)G_{\text{SM}}}{ \mathcal{N}_{\text{M}}}.
\end{equation}

\section{Conditions for Full Duplex Gains}\label{sec:SB_Comp}
In this section, we will discuss the capacity of FD operations, and compare it with the HD counterparts to derive conditions favorable for FD operation. Note that for the capacity comparison in this section, we assume a  bufferless  small cell BS, unlike Section~\ref{sec:SB_sched} where the availability of buffering at the small cell BS will be considered. In a HD system, the joint spectral efficiency is given by,
\begin{equation}\label{SB_joint_cap_HD}
\begin{split}
\text{C}_{\text{HD}} =&\underbrace{0.5~\text{log}_2 (1+ \text{min} (\text{SNR}_{\text{S}}^{\text{HDD}}, \text{SNR}_{\text{D}}^{\text{HDD}} ))}_{\text{downlink}} \\
&+ \underbrace{0.5~\text{log}_2 (1+\text{min} (\text{SNR}_{\text{S}}^{\text{HDU}}, \text{SNR}_{\text{M}}^{\text{HDU}}))}_{\text{uplink}},
\end{split}
\end{equation}
where factor 0.5 comes from the fact that the channel is equally divided between uplink and downlink. 

In case of FD transmissions, let us divide all the four transmission modes, i.e, FDD, FDU, FDB, and FDA into two joint modes, (1) FD mode~1, which includes FDD and FDU modes, (2) FD mode~2, which includes FDB and FDA modes. The joint spectral efficiency of FD mode~1 can be defined as
\begin{equation}\label{SB_joint_cap_FDM1}
\begin{split}
\text{C}_{\text{FD}}^{\text{Mode1}} =&\underbrace{\text{log}_2 (1+ \text{min} (\text{SINR}_{\text{S}}^{\text{FDD}}, \text{SINR}_{\text{D}}^{\text{FDD}} ))}_{\text{downlink}} \\
&+ \underbrace{\text{log}_2 (1+\text{min} (\text{SINR}_{\text{S}}^{\text{FDU}}, \text{SINR}_{\text{M}}^{\text{FDU}}))}_{\text{uplink}}.
\end{split}
\end{equation}

Similarly, the joint spectral efficiency of FD mode~2 is given by
\begin{equation}\label{SB_joint_cap_FDM2}
\begin{split}
\text{C}_{\text{FD}}^{\text{Mode2}} =&\underbrace{\text{log}_2 (1+ \text{min} (\text{SINR}_{\text{S}}^{\text{FDB}}, \text{SINR}_{\text{D}}^{\text{FDA}} ))}_{\text{downlink}} \\
&+ \underbrace{\text{log}_2 (1+\text{min} (\text{SINR}_{\text{S}}^{\text{FDA}}, \text{SINR}_{\text{M}}^{\text{FDB}}))}_{\text{uplink}}.
\end{split}
\end{equation}

\begin{figure}
\centering
\includegraphics[width = 2.3 in] {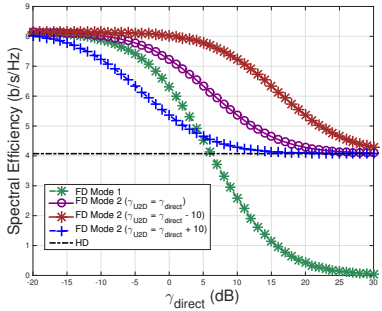}
\vspace{-2mm}
\caption{Instantaneous spectral efficiency of the HD and FD modes with $\gamma_{\text{SI}} = 0$ but with different values of $\gamma_{\text{direct}}$. For other links, $10\text{log}_{10}(\frac{G_{\text{MS}}}{\mathcal{N}_{\text{S}}}) =10\text{log}_{10} (\frac{G_{\text{SM}}}{\mathcal{N}_{\text{M}}}) = 10\text{log}_{10} (\frac{G_{\text{SD}}}{\mathcal{N}_{\text{D}}}) = 10\text{log}_{10} (\frac{G_{\text{US}}}{\mathcal{N}_{\text{S}}}) = 12$~dB. }
\label{fig:SB_comp_noSI}
\vspace{-6mm}
\end{figure}

With these expressions, we compare the performance of different modes under different interference conditions. In this section we use fixed power allocation and set the values of powers as $p_{\text{M}}(t) = p_{\text{S}}(t) = p_{\text{U}}(t) = 1 $ watt. Fig.~\ref{fig:SB_comp_noSI} shows the spectral efficiency of different modes when there is no self-interference. In this figure, we assume the values of all the single hop channel-SNR equal to 12 dB, i.e, $10\text{log}_{10}(\frac{G_{\text{MS}}}{\mathcal{N}_{\text{S}}}) =10\text{log}_{10} (\frac{G_{\text{SM}}}{\mathcal{N}_{\text{M}}}) = 10\text{log}_{10} (\frac{G_{\text{SD}}}{\mathcal{N}_{\text{D}}}) = 10\text{log}_{10} (\frac{G_{\text{US}}}{\mathcal{N}_{\text{S}}}) = 12$~dB. Spectral efficiency of FD mode~1 is shown with the different strengths of the links between macro BS and UEs, for which we consider $10\text{log}_{10}(\frac{G_{\text{MD}}}{\mathcal{N}_{\text{D}}}) = 10\text{log}_{10}(\frac{G_{\text{UM}}}{\mathcal{N}_{\text{M}}}) = \gamma_{\text{direct}}$. The spectral efficiency of FD mode~2 is shown for the different values of the interference channel between the downlink UE and the uplink UE, i.e., $10\text{log}_{10}(\frac{G_{\text{UD}}}{\mathcal{N}_{\text{D}}}) = \gamma_{\text{U2D}}$.

In FD mode~1, the performance of the downlink and uplink transmissions are affected by the interference at the downlink UE and at the macro BS, respectively. The higher values of $\gamma_{\text{direct}}$ lowers the spectral efficiency of the FD mode~1, even below the HD mode. In  FD mode~2, when there is no self-interference, only the downlink UE experiences interference which comes from the uplink UE. We consider three cases to compare the performance of FD mode~1 and FD mode~2. In the cases, when $\gamma_{\text{U2D}}$ is lower or equal to $\gamma_{\text{direct}}$, FD mode~2 performs better than the FD mode~1. In the case when $\gamma_{\text{U2D}}$ is higher than the $\gamma_{\text{direct}}$, for the lower values of $\gamma_{\text{direct}}$, FD mode~1 performs better than the FD mode~2, but after a certain point, the performance of FD mode~1 becomes worse. This is because in FD mode~1, increasing $\gamma_{\text{direct}}$ lowers both the uplink and downlink SINRs, whereas in FD mode~2, uplink SINR is not affected by any interference, only the downlink SINR is decreased by increasing $\gamma_{\text{U2D}}$. Moreover, since in this case we consider the symmetric channel gain for the downlink and uplink access links, i.e., $G_{\text{SD}} = G_{\text{US}}$, which makes the uplink spectral efficiency of FD mode~2 equal to the joint spectral efficiency of the HD mode, i.e., $\text{log}_2 (1+\text{min} (\text{SINR}_{\text{S}}^{\text{FDA}}, \text{SINR}_{\text{M}}^{\text{FDB}})) = \text{C}_{\text{HD}}$, which always provides higher spectral efficiency in FD mode~2 than the HD mode, even if the downlink spectral efficiency becomes lower.  
\begin{figure}
\centering
\includegraphics[width = 2.3 in] {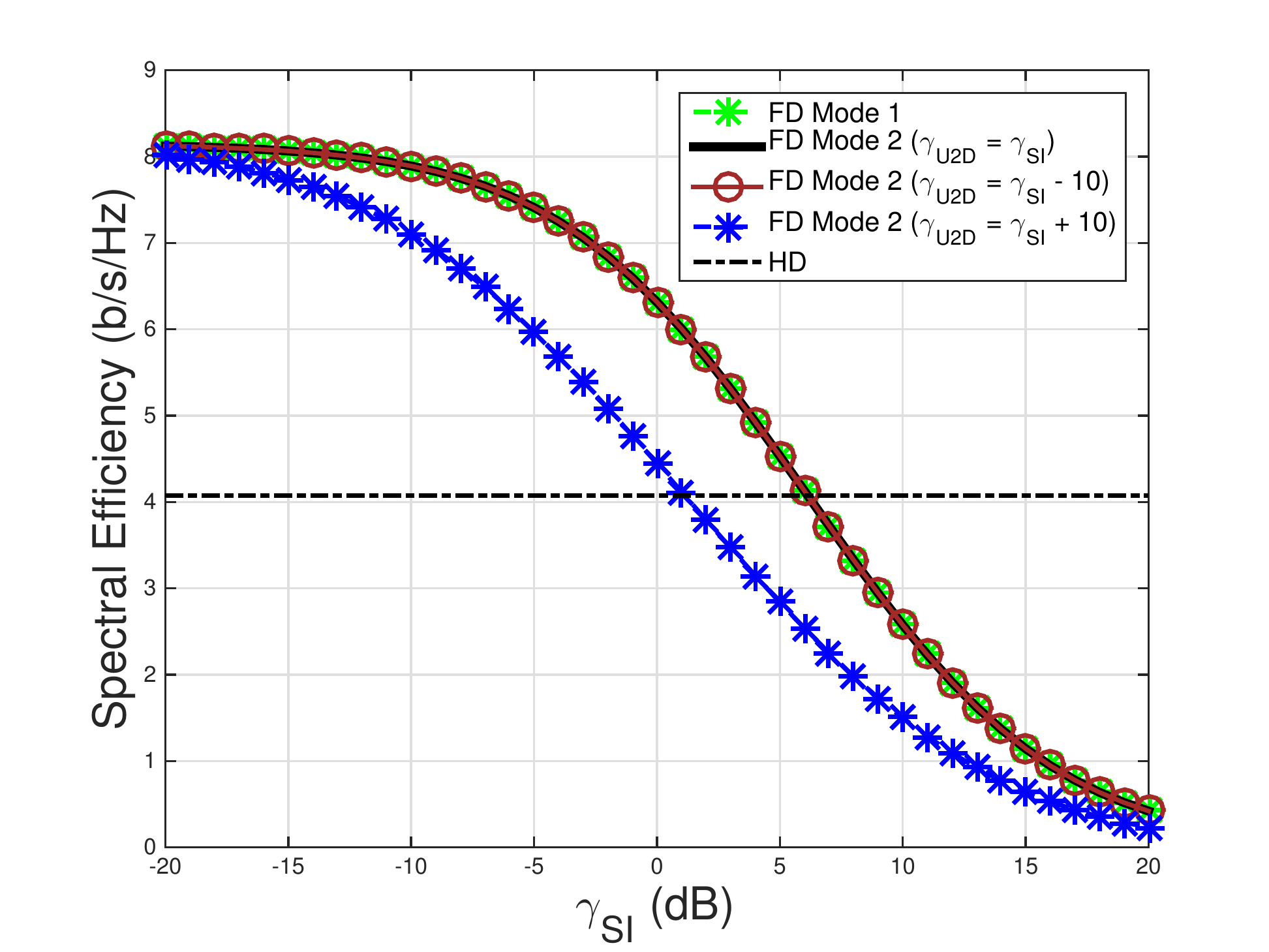}
\vspace{-2mm}
\caption{Instantaneous spectral efficiency of the HD and FD modes $\gamma_{\text{direct}} = 0$ but with different values of $\gamma_{\text{SI}}$. For other links, $10\text{log}_{10}(\frac{G_{\text{MS}}}{\mathcal{N}_{\text{S}}}) =10\text{log}_{10} (\frac{G_{\text{SM}}}{\mathcal{N}_{\text{M}}}) = 10\text{log}_{10} (\frac{G_{\text{SD}}}{\mathcal{N}_{\text{D}}}) = 10\text{log}_{10} (\frac{G_{\text{US}}}{\mathcal{N}_{\text{S}}}) = 12$~dB.}
\label{fig:SB_comp_noSD}
\vspace{-6mm}
\end{figure}

In Fig.~\ref{fig:SB_comp_noSD}, we neglect the interference between the macro BS and the UEs, i.e., $\gamma_{\text{direct}} = 0$ (in linear scale) and vary self-interference at both macro BS and small cell BS, for which we assume, i.e., $10\text{log}_{10}(\frac{\gamma_{\text{S}}}{\mathcal{N}_{\text{S}}}) = 10\text{log}_{10}(\frac{\gamma_{\text{M}}}{\mathcal{N}_{\text{M}}}) = \gamma_{\text{SI}}$. In this case, self-interference $(\gamma_{\text{SI}})$ decreases the SINRs of both downlink and uplink transmissions in both FD mode~1 and FD mode~2. Thus, as we increase the $\gamma_{\text{SI}}$, as shown in Fig.~\ref{fig:SB_comp_noSD}, the spectral efficiency of both FD mode~1 and FD mode~2 decrease. They become lower than the HD spectral efficiency after some point. Moreover, in case of FD mode~2, the downlink transmission is also affected by the interference from the uplink UE ($\gamma_{\text{U2D}}$). In the cases when the interference from the uplink UE is higher than the self-interference, i.e., $\gamma_{\text{U2D}}$ > $\gamma_{\text{SI}}$, the dominating interference in downlink is interference from the uplink UE which decreases the spectral efficiency of FD mode~2 lower than FD mode~1. In other cases the dominating interference in downlink is the self-interference, so it provides the same spectral efficiency as  FD mode~2. 

The above results show that the performance of a transmission mode depends on the different channel conditions. A transmission mode should be chosen depending on its favorable channel condition to receive the maximum spectral efficiency gain, which is the topic of the next section. 

\section{Joint Scheduling and Power Allocation}\label{sec:SB_sched}
In this section we consider a problem of joint scheduling and power allocation. In each time slot, the scheduler can schedule different links as shown in the Section~\ref{sec:SB_SM}. In our system with a FD macro BS and a FD small BS, we allow both FD and HD transmissions. Thus, in each time slot, the scheduler can either select one of FD transmission modes, i.e., FDD, FDU, FDB, and FDA modes or can select one of HD modes shown in Fig.1. A transmission mode is selected based on the channel conditions such that the maximum capacity gain can be achieved. Moreover, since in general there are multiple UE traffic flows, appropriate traffic flow should be scheduled on each link so that all the queues corresponding to all the UEs remain stable. This problem is equivalent to a scheduling problem in multi-hop wireless networks, where the back-pressure based scheduling given by Tassiulas~\cite{tassiluas_bp} is a well-known throughput-optimal algorithm. We apply the same back-pressure based algorithm in this case, but in addition an optimal power allocation is applied to minimize the interference during the FD transmission modes. During the HD transmission modes, since there is no interference, selected nodes transmit at maximum power to achieve the maximum throughput. 

To define our scheduling algorithm, we first assign a weight to each link. Please note that the downlink and uplink link between any two nodes are considered as separate links. At the beginning of time slot $t$, each link $l$ is assigned a link weight equal to the maximum backlog differential of all the flows passing through the link:
\begin{equation}\label{SB_link_wt}
\mathrm{W}_{l}(t) = \mathrm{\max}_{n \in \{1,2,\cdots,N\}} \left( Q^n_{l_i}(t) - Q^n_{l_j}(t)\right),
\end{equation}
\begin{equation}\label{SB_link_wt}
l_{f}(t) = \argmax_{n \in \{1,2,\cdots,N\}} \left( Q^n_{l_i}(t) - Q^n_{l_j}(t)\right),
\end{equation}
where $Q^n_{l_i}(t)$ and $Q^n_{l_j}(t)$ are the queue backlog corresponding to UE $n$ on the source node of the link $l$ ($l_i$) and the destination node of the link $l$ ($l_j$), respectively, at time $t$. In this formulation, a node can represent the macro BS, the small BS, and any of the UEs. Moreover, in our system, a UE does not forward data for other UEs, so there will no link between two UEs. There will also be no direct transmission between the macro BS and UE. If the link $l$ is selected by the scheduler, then packets belonging to UE $l_{f}(t)$ will be transmitted.

After assigning the weight to each link, a schedule $\pi(t)$ is derived such that
\begin{equation}\label{SB_pf}
\pi(t) = \argmax_{\tau \in \Gamma} \sum_{l \in \tau} W_l(t) R^*_l(\tau,t),
\end{equation}
where $\Gamma$ is the set of all feasible schedules, which in our case, consists all transmission modes as shown in Figs.~\ref{fig:SB_diff_FDM} and~\ref{fig:SB_diff_HDM}. In case of FD modes, $\tau$ will contain two links to schedule simultaneously. $R^*_l(\tau,t)$ is the data rate on link $l$ selected in schedule $\tau$. In case of FD modes, the data rate on each link will include the interference from the other link scheduled simultaneously, while using the optimal power allocation derived for each link, the details of which is given following in the next subsection. 

\subsection{Power Allocation}\label{sec:SB_PC}
Given the channel gains, in each of the FD modes, we find the optimal transmit power for each node such that the weighted sum rate of both the links is maximized. The weights of the links are derived from~(\ref{SB_link_wt}). Let us consider the FDD mode, where the SINR at the small BS and a downlink UE D are given in~(\ref{SB_sinr_BS_D_FDD}). In this case, the power allocation problem can be written as
\begin{equation}\label{SB_pc_FDD}
\begin{aligned}
\{p^*_{\text{M}}(t), p^*_{\text{S}}(t) \} = \argmax_{\substack {p_{\text{M}} \in [0, p_{\text{M}}^{max}] \\  p_{\text{S}} \in [0, p_{\text{S}}^{max}] }} \Bigg[ W_{\text{MS}}(t)~\mathrm{log}\left(1+ \frac{p_{\text{M}} G_{\text{MS}}}{ p_{\text{S}} \gamma_{\text{S}} + \mathcal{N}_{\text{S}}}\right)\\
+~W_{\text{SD}}(t)~\mathrm{log} \left(1 + \frac{p_{\text{S}} G_{\text{SD}}}{ p_{\text{M}} G_{\text{MD}} + \mathcal{N}_{\text{D}}} \right)\Bigg], 
\end{aligned}
\end{equation}
where $W_{\text{MS}}(t)$ and $W_{\text{SD}}(t)$ are the weights of the link between macro BS and small BS, and the link between small BS and downlink UE D, respectively.

The above optimization~(\ref{SB_pc_FDD}) is a nonlinear nonconvex problem. We use Geometric Programming (GP)~\cite{boyd2007tutorial,chiang2007power} to get a near-optimal solution of~(\ref{SB_pc_FDD}). The problem~(\ref{SB_pc_FDD}) can be written as
\begin{equation}\label{SB_pc_GP_stan}
\footnotesize{
\begin{aligned}
&\{p^*_{\text{M}}(t), p^*_{\text{S}}(t) \} = \argmin_{\{x,y\}} \Bigg[\left( \frac{y \gamma_{\text{S}} + \mathcal{N}_{\text{S}}}{x G_{\text{MS}} + y \gamma_{\text{S}} + \mathcal{N}_{\text{S}}}\right)^{W_{\text{MS}}(t)} \\
 &~~~~~~~~~~~~~+ \left( \frac{x G_{\text{MD}} + \mathcal{N}_{\text{D}}}{y G_{\text{SD}} + x G_{\text{MD}} + \mathcal{N}_{\text{D}}}\right)^{W_{\text{SD}}(t)} \Bigg] \\
  &\mbox{subject to:} \\
    & \ \ \ \ \ \ \ \ 0 \le \frac{x}{p_{\text{M}}^{max}} \le 1, 0 \le \frac{y}{p_{\text{S}}^{max}} \le 1 .
\end{aligned}
}
\end{equation}

In general, to apply GP, the optimization problem should be in GP standard form~\cite{boyd2007tutorial, chiang2007power}. In the GP standard form, the objective function is a minimization of a $\textit {posynomial}$\footnote{ A monomial is a function $f:\mathbf{R}_{++}^n \rightarrow \mathbf{R}: g(p) = d p_1^{a^{(1)}}p_2^{a^{(2)}}\cdots p_n^{a^{(n)}}$, where $d \geq 0$ and $a^{(k)} \in \mathbf{R}, k = 1,2,\cdots,n.$ A posynomial is a sum of monomials, $f(p) = \sum_{j=1}^J d_j p_1^{a_j^{(1)}} p_2^{a_j^{(2)}} \cdots p_n^{a_j^{(n)}}$. } function; the inequalities and equalities in the constraint set are a posynomial upper bound inequality and $\textit {monomial}$ equality, respectively. 

In our case, in (\ref{SB_pc_GP_stan}), constraints are monomials (hence posynomials), but the objective function is a ratio of posynomials. Hence, (\ref{SB_pc_GP_stan}) is not a GP in standard form, because posynomials are closed under multiplication and addition, but not under division.

According to~\cite{chiang2007power}, (\ref{SB_pc_GP_stan}) is a signomial programming (SP) problem. In~\cite{chiang2007power}, an iterative procedure is given, in which (\ref{SB_pc_GP_stan}) is solved by constructing a series of GPs, each of which can easily be solved. In each iteration of the series, the GP is constructed by approximating the denominator posynomial (\ref{SB_pc_GP_stan}) by a monomial, then using the arithmetic-geometric mean inequality and the value of $\{x,y\}$ from the previous iteration. The series is initialized by any feasible $\{x,y\}$, and the iteration is terminated at the $s_{th}$ loop if $||{x}_s - x_{s-1}|| < \epsilon $, and $||{y}_s - y_{s-1}|| < \epsilon $, where $\epsilon$ is the error tolerance. This procedure is provably convergent, and empirically almost always computes the optimal power allocation~\cite{chiang2007power}.  

Finally, the derived values of the powers ($p^*_{\text{M}}(t), p^*_{\text{S}}(t)$) are used to calculate the optimal rates of the links as
\begin{equation}\label{SB_pc_Rates}
\begin{aligned}
R^*_{\text{MS}} (t) = \mathrm{log}\left(1+ \frac{p^*_{\text{M}}(t) G_{\text{MS}}}{ p^*_{\text{S}}(t)\gamma_{\text{S}} + \mathcal{N}_{\text{S}}}\right),\\
R^*_{\text{SD}} (t) = \mathrm{log} \left(1 + \frac{p^*_{\text{S}}(t) G_{\text{SD}}}{ p^*_{\text{M}}(t) G_{\text{MD}} + \mathcal{N}_{\text{D}}} \right) 
\end{aligned}
\end{equation}

These rates are used in the scheduling decision~(\ref{SB_pf}), where the optimal rates using the above power allocation method are calculated for all possible combination of link schedules. A similar method is used to derive the optimal rates in the other FD transmission modes.

Applying the power allocation method for each possible FD transmission mode in each time slot adds high computation complexity, which increases exponentially with the number of UEs. To solve this time complexity problem, we use a sub-optimal scheduling method. For each possible FD transmission consisting all combinations in FDD, FDU, FDB, and FDA modes,  we initially assign equal weight to all the links (e.g., $W_l = 1, \forall l$) for all the UEs, assuming equal traffic demand for all, and determine the power allocation using the method described above. Then in each time slot, first, with the initial power allocation, a scheduling decision is derived in each of the four FD modes. Then in each mode, the optimal power allocation is derived only for the chosen scheduling decision with the method described above. Finally, the chosen scheduling decisions with optimal power allocation in each mode are compared with each other, and also with all the HD modes to find the best possible scheduling decision to schedule as given in~(\ref{SB_pf}).

\section{Performance Evaluation}\label{sec:SB_PE}
We evaluate the performance of the proposed hybrid dynamic scheduling (\emph{FD System}), and make a comparison with the baseline system where only HD transmissions are allowed (\emph{HD System}). We also show the effect of power allocation in the FD transmissions described in Section~\ref{sec:SB_PC}. As described in Section~\ref{sec:SB_SM}, we consider a macro-cell with a macro BS at the center of the cell and an outdoor small cell serving 10 UEs. Other simulation parameters are listed in Table~\ref{tab:SB_sim_par}, and are based on 3GPP simulation recommendations for outdoor environments~\cite{3GPP36828}. The probability of LOS for all the channels can be found in \cite[Table 6.4-1]{3GPP36828}. We capped the spectral efficiency at 7 bits/sec/Hz to match the peak spectrum efficiency of a system with practical modulation and coding. We assume 120 dB of self-interference cancellation at both macro and small BSs during the FD operations.  

\begin{table}
\centering
{\scriptsize
\caption {Simulation parameters. Here, SSD: Shadowing standard deviation, MBS: macro BS, SBS: small BS, and R is in kilometers.} \label{tab:SB_sim_par} 
 \vspace{-1mm}
\begin{tabular}{| p{1.1 in} | p{1.8 in} |}
\hline
\textbf {Parameter} & \textbf{Value} \\ \hline
System bandwidth & 10 MHz \\ \hline
		Radius of macro-cell & 800 m \\ \hline
		Radius of a small cell & 40 m \\ \hline
		Maximum power & MBS: 46 dBm, SBS: 24 dBm, UE: 23 dBm \\ \hline
		Noise figure & MBS: 5 dB, SBS: 13 dB, UE: 9 dB \\ \hline
		
		SSD between SBS and UE &  LOS:~ 3 dB NLOS: 4 dB \\  \hline
		SSD between MBS and UE &  8 dB \\  \hline
		SSD between MBS and SBS &  6 dB \\  \hline
		MBS to SBS path loss  & LOS: $PL(R) = 100.7 + 23.5~log_{10}(R),$ NLOS: $PL(R) = 125.2 + 36.3~log_{10}(R)$.  \\ \hline
		MBS to UE path loss  & LOS: $PL(R) = 103.4 + 24.2~log_{10}(R),$ NLOS: $PL(R) = 131.1 + 42.8~log_{10}(R)$.  \\ \hline
	         SBS to UE path loss  & LOS: $PL(R) = 103.8 + 20.9~log_{10}(R),$ NLOS: $PL(R) = 145.4 + 37.5~log_{10}(R)$.  \\ \hline
		UE to UE path loss & If $R \leq 50 m, PL(R) = 98.45 + 20~log_{10}(R),$ else,  $PL(R) = 55.78 + 40~log_{10}(R)$.  \\ \hline
	\end{tabular}
	}
	 \vspace{-4mm}
\end{table}

The scenario we simulated is a fixed macro BS with several randomly located small cell BS within the coverage area of the macro BS. For each location of small cell BS, UEs are randomly dropped in the small cell.  We evaluated the system with a practical FTP traffic model recommended by 3GPP~\cite{3GPP36814}, where each UE generates requests to download and  upload files. The time interval between completion of a file transmission and an arrival of a new request is exponentially distributed with a mean of 1 second.

\begin{figure}[t!]
\begin{subfigure}[t]{0.22\textwidth}
        \centering
        \includegraphics[width = 1.8 in] {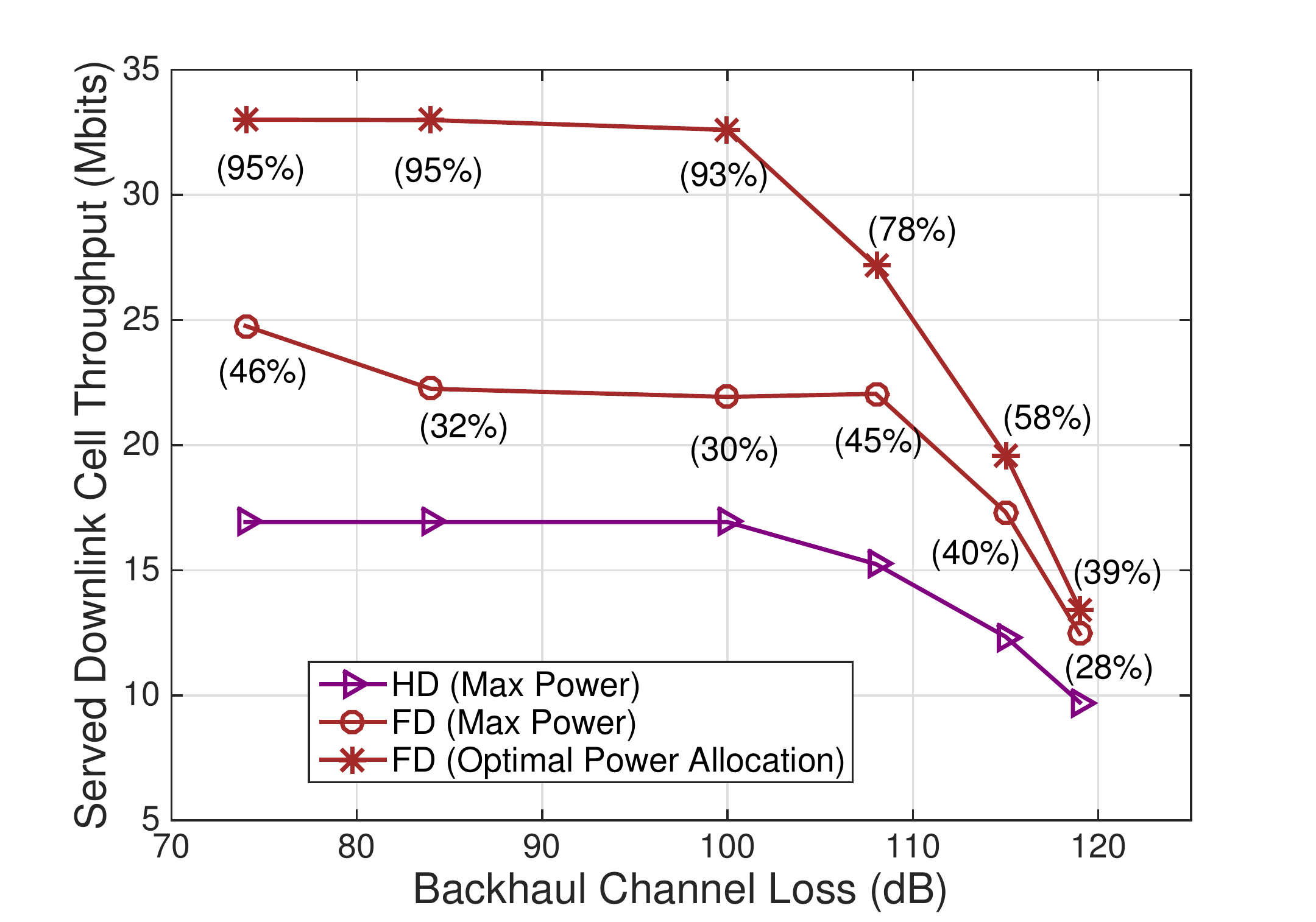}
	\vspace{-4mm}
	\caption{Downlink}
	\label{fig:SB_down_servTh_eqDeman}
        \end{subfigure}%
        ~ 
    \begin{subfigure}[t]{0.22\textwidth}
        \centering
        \includegraphics[width = 1.8 in] {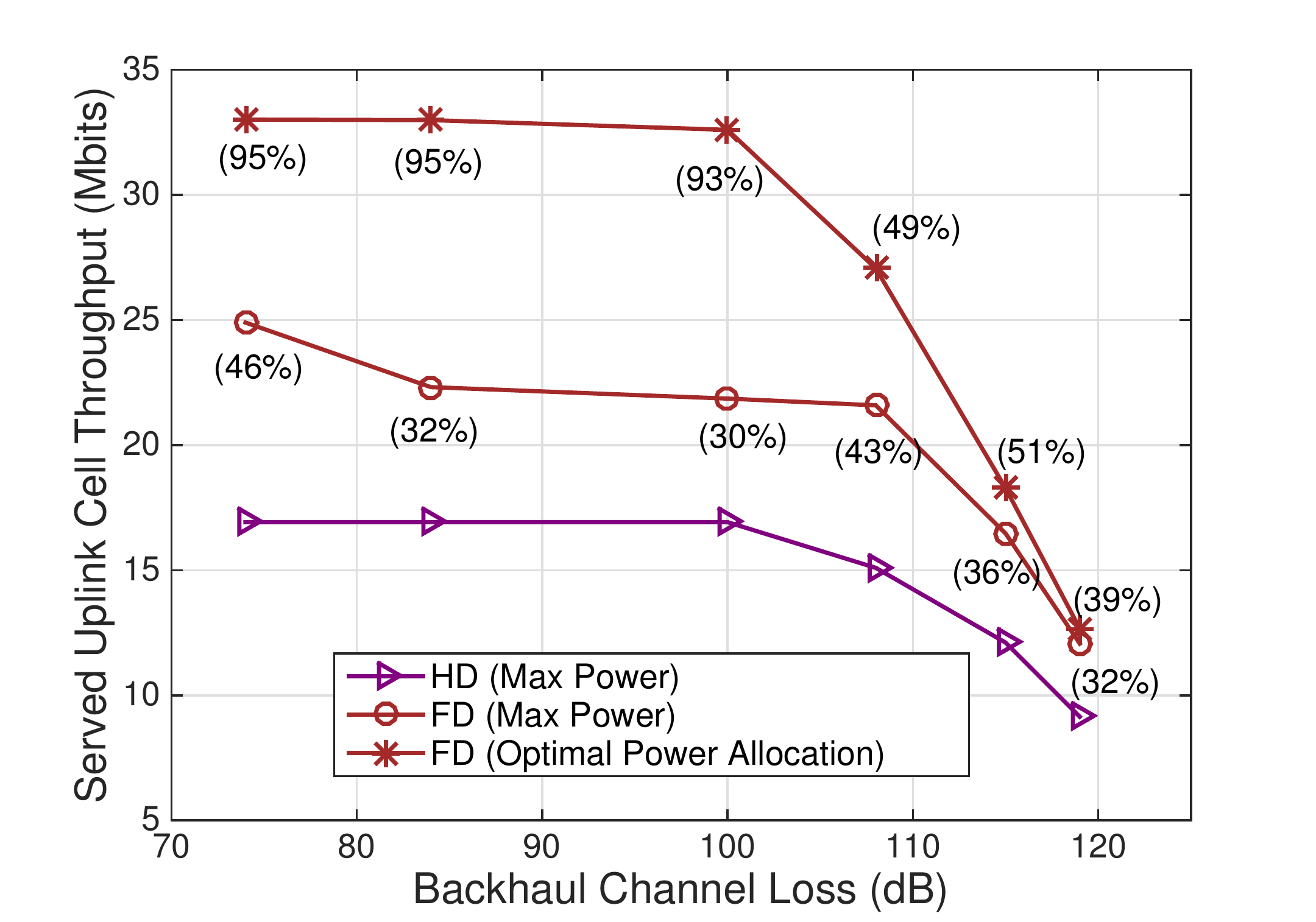}
	\vspace{-4mm}
	\caption{Uplink}
	\label{fig:SB_up_servTh_eqDeman}
    \end{subfigure}
    \vspace{-2mm}
    \caption{Served cell throughput with equal uplink and downlink traffic demands.}
     \vspace{-6mm}
\end{figure}

We first simulate the case with equal uplink and downlink demands in which each UE  uploads and downloads files of the same size of 1.25 MB. For each instance of small BS and its UE locations, our simulation runs for 50 secs. Figures~\ref{fig:SB_down_servTh_eqDeman} and~\ref{fig:SB_up_servTh_eqDeman} show the served cell throughput (defined by the total amount of data served for all users over the total amount of observation time) in the downlink and uplink, respectively. For the FD system, the results are plotted for two cases: one with the proposed power allocation in Section~\ref{sec:SB_PC}, and the other with fixed power at maximum level. In the figures, the percentages in brackets represent the gains compared to the HD system. We can see both FD systems have significant throughput gains compared to the HD system, while the optimal power allocation has significant gains over fixed power allocation. This demonstrates the importance of interference management for the FD operations. For all the systems, throughput decreases as the backhaul channel gain decreases. In the case of a strong backhaul channel, where the small BS and macro BS are close to each other, interfering links between macro BS and uplink/downlink UEs are strong. However, in our hybrid scheduling method, where mixing FD/HD modes are allowed, the scheduler makes sure strong interference is avoided. In addition to that, our power allocation method adjusts powers of the transmitters, which further minimizes the interference and provides higher FD gains. In the case with a weaker backhaul channel, where the small cell BS is far from the macro-cell BS, the backhaul link becomes  the bottleneck. Both hybrid scheduling and power allocation do not provide much improvement, Therefore, FD gain decreases as the backhaul channel strength decreases.  

\begin{table}
\centering
{\scriptsize
\caption {Average number of transmissions in different modes with different backhaul channel loss (BCL). Here FP: Fixed (Max) Power, and PA: Power Allocation.} 
 \vspace{-1mm}\label{tab:SB_no_tx_equaldeman} 
\begin{tabular}{|c|c|c|c|c|c|c|}
\hline
{} & \begin{tabular}[c]{@{}c@{}} \textbf{BCL: 74 dB } \\(FP, PA)    \end{tabular} & \begin{tabular}[c]{@{}c@{}} \textbf{BCL: 100 dB } \\(FP, PA)    \end{tabular} & \begin{tabular}[c]{@{}c@{}} \textbf{BCL: 119 dB } \\(FP, PA)    \end{tabular} \\ \hline
 \textbf{HD}            & (27\%, 3\%)                        & (35\%, 3\%)  & (63.6\%, 56.6\%)  \\ \hline
  \textbf{FDD}            & (1\%, 33\%)                        & (32\%, 47\%)  & (18\%, 24\%)  \\ \hline
 \textbf{FDU}            & (4\%, 33\%)                       & (23\%, 47\%)  & (18\%, 19\%)  \\ \hline
 \textbf{FDB}            & (35\%, 15\%)                   & (1\%, 2\%)  & (0.2\%, 0.2\%)  \\ \hline
 \textbf{FDA}            & (33\%, 16\%)                  & (9\%, 1\%)  & (0.2\%, 0.2\%)  \\ \hline

\end{tabular}
} \vspace{-4mm}
\end{table}

We also collected statistics about how frequent each mode is used for all our simulated scenarios, and the results are shown in Table~\ref{tab:SB_no_tx_equaldeman}. With the proposed power allocation method, the proposed scheduler is able to tune the transmission power to an appropriate levels, which allows the system to transmit in FD mode more often when compared with a system that has fixed power level. For example, in the fixed power scenario, when the backhaul channel is strong, FDD and FDU modes are rarely scheduled. This is due to the strong interference between macro BS and UEs. With the proposed power allocation method, 66\% transmissions are  scheduled in FDD and FDU modes. When the backhaul channel is weak, most of the transmissions are scheduled in HD modes, since most of transmissions are scheduled in the downlink backhaul link. This is because when the backhaul capacity is low,  the backhaul link needs to be scheduled  more frequently to match the access link capacity.  
\begin{figure}[t!]
\begin{subfigure}[t]{0.22\textwidth}
        \centering
        \includegraphics[width = 1.8 in] {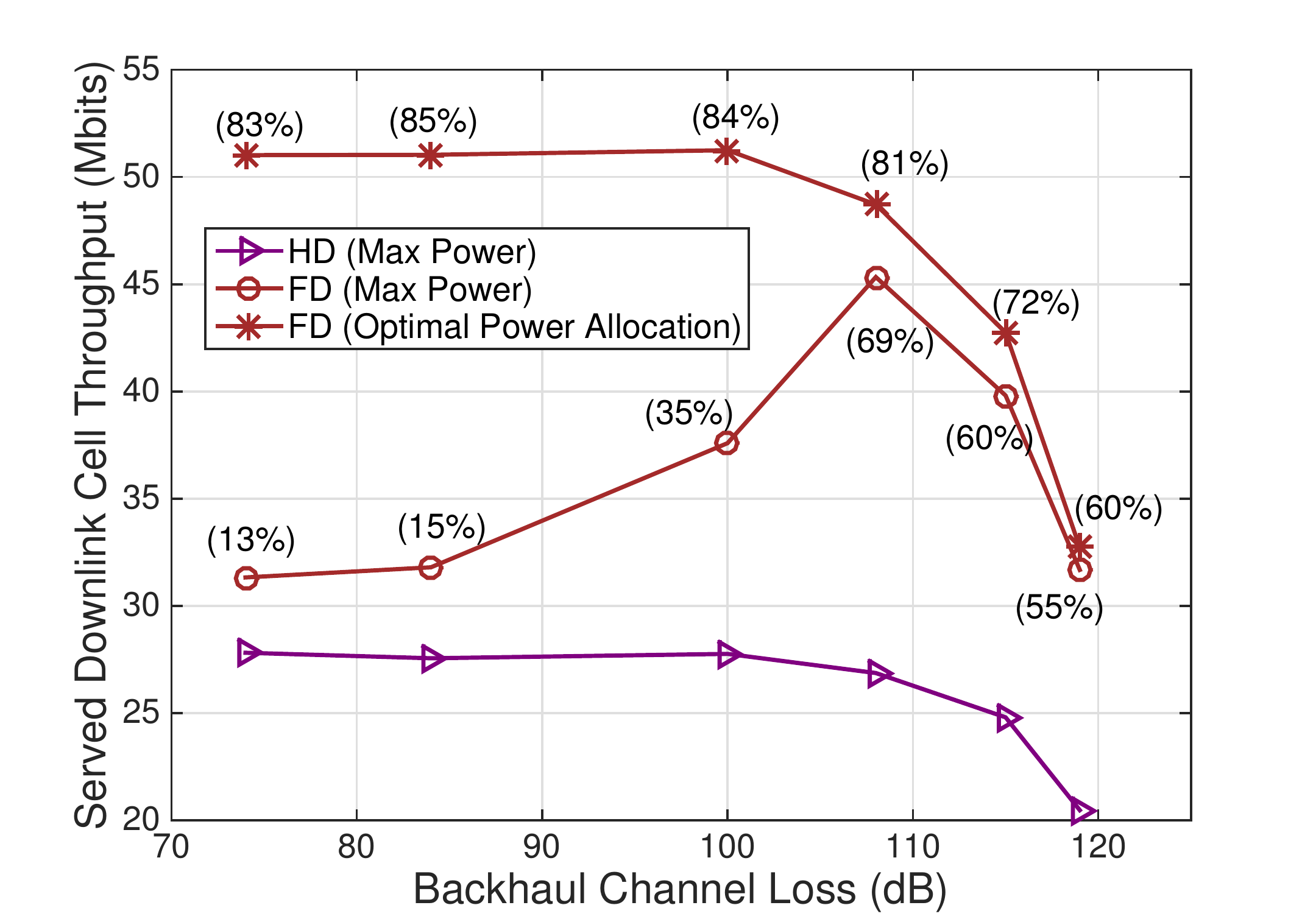}
	\vspace{-4mm}
	\caption{Downlink}
	\label{fig:SB_down_servTh_uneqDeman}
        \end{subfigure}%
        ~ 
    \begin{subfigure}[t]{0.22\textwidth}
        \centering
        \includegraphics[width = 1.8 in] {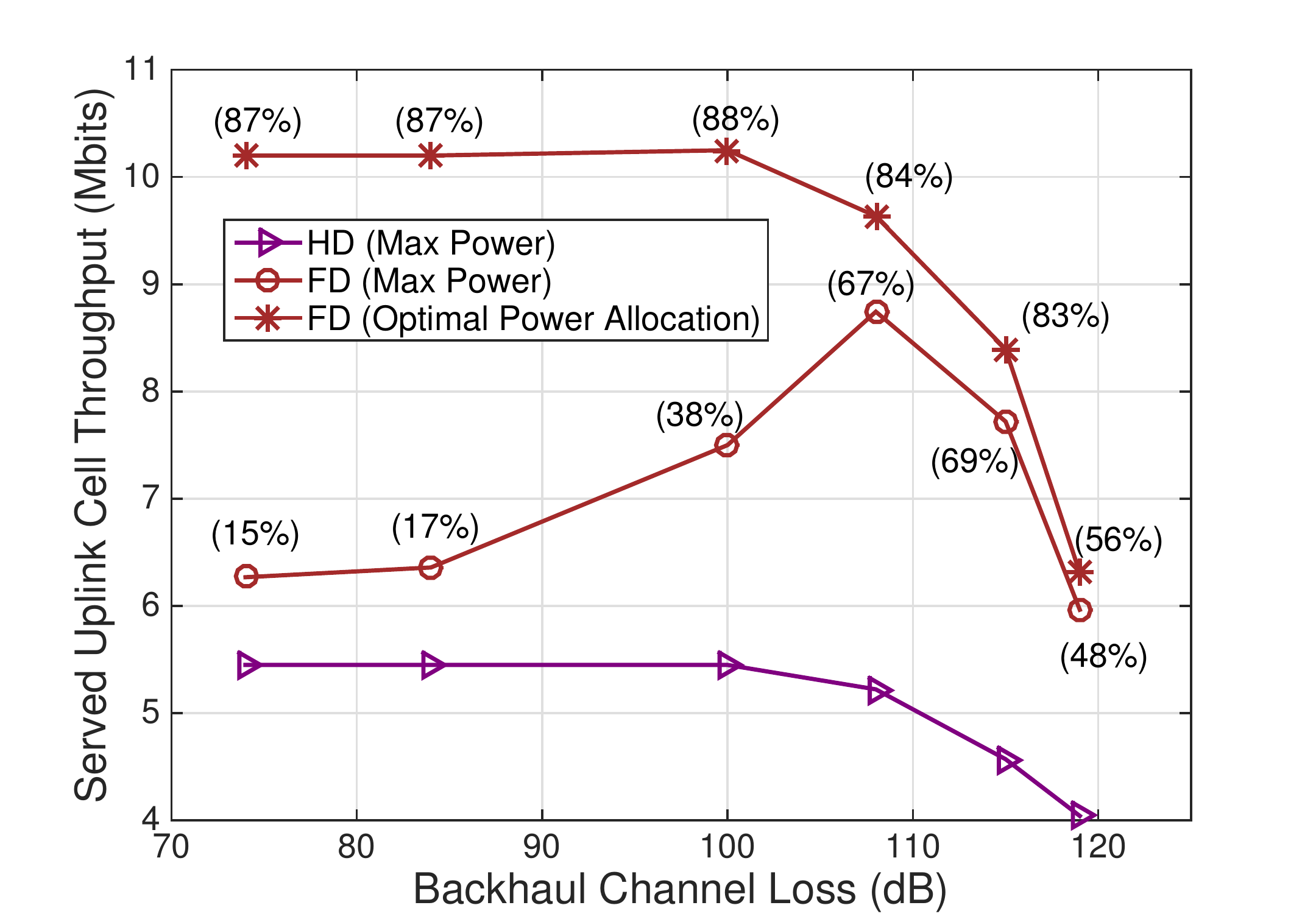}
	\vspace{-4mm}
	\caption{Uplink}
	\label{fig:SB_up_servTh_uneqDeman}
    \end{subfigure}
    \vspace{-2mm}
    \caption{Served cell throughput when the downlink traffic demand is fives times more than the uplink traffic demand.}
     \vspace{-6mm}
\end{figure}

To simulate a system with asymmetric downlink and uplink traffic demand, we consider a case where downlink traffic demand is five times larger than the uplink traffic demand for each UE. 
%Since during the scheduling decisions, our scheduler includes the traffic demand in its decision, it can tune its decisions according to the required traffic demand. 
All the other simulation parameters are the same as the previous case of equal demands. Each UE uploads and downloads files of size 250KB and 1.25 MB, respectively. The results are shown in Figures~\ref{fig:SB_down_servTh_uneqDeman} and~\ref{fig:SB_up_servTh_uneqDeman}. In the unequal traffic demand scenario, the FD system still achieves significant throughput gains compared to the HD system. However, as we can see, the uplink throughput is close to one fifth of the downlink throughput, which matches the traffic demands of the FTP application. In the case of fixed power assignment with strong backhaul channel, throughput gains are lower when  compared to the previous results for symmetric traffic demands. As explained earlier, for the case of strong backhaul channel, interference between the macro-cell and the UEs are also strong. Therefore, without any power optimization, FDD and FDU modes are rarely selected due to the strong interference. Moreover, in this case of asymmetric traffic demands, opportunities for selecting the FDB and FDA modes are also low, therefore only HD modes are scheduled. This is the reason why lower throughput gains are achieved with fixed power allocation, when the small BS is near to the macro BS. 
FD gain increases as the backhaul channel starts to become weaker. However, after a certain point, the backhaul channel becomes the bottleneck. FD transmission opportunities become harder to find and the gain starts to drop, which was also seen in the previous case. 

\begin{figure}[t!]
\begin{subfigure}[t]{0.22\textwidth}
        \centering
        \includegraphics[width = 1.8 in] {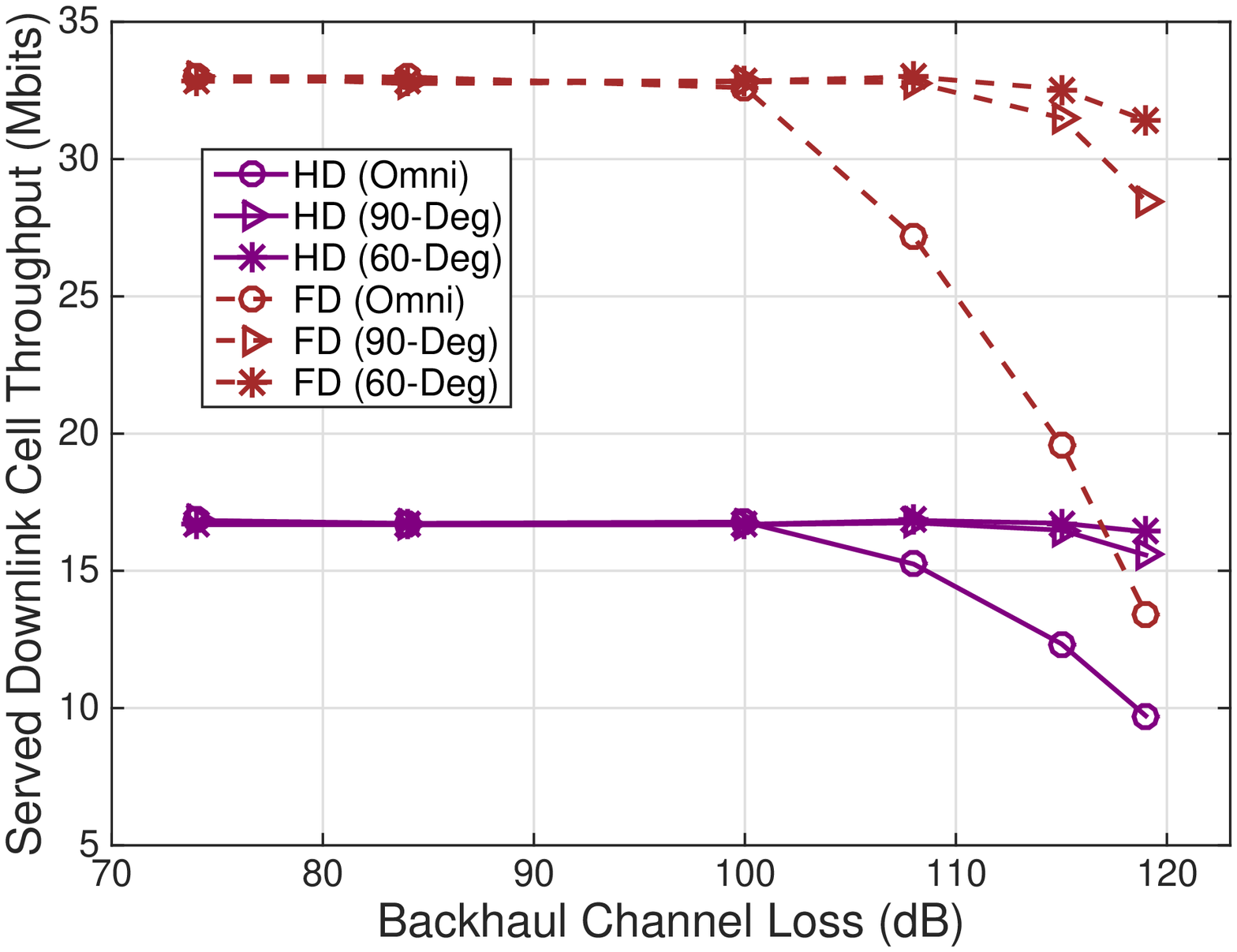}
	\vspace{-4mm}
	\caption{Downlink}
	\label{fig:SB_down_servTh_dir_eqDeman}
        \end{subfigure}%
        ~ 
    \begin{subfigure}[t]{0.22\textwidth}
        \centering
        \includegraphics[width = 1.8 in] {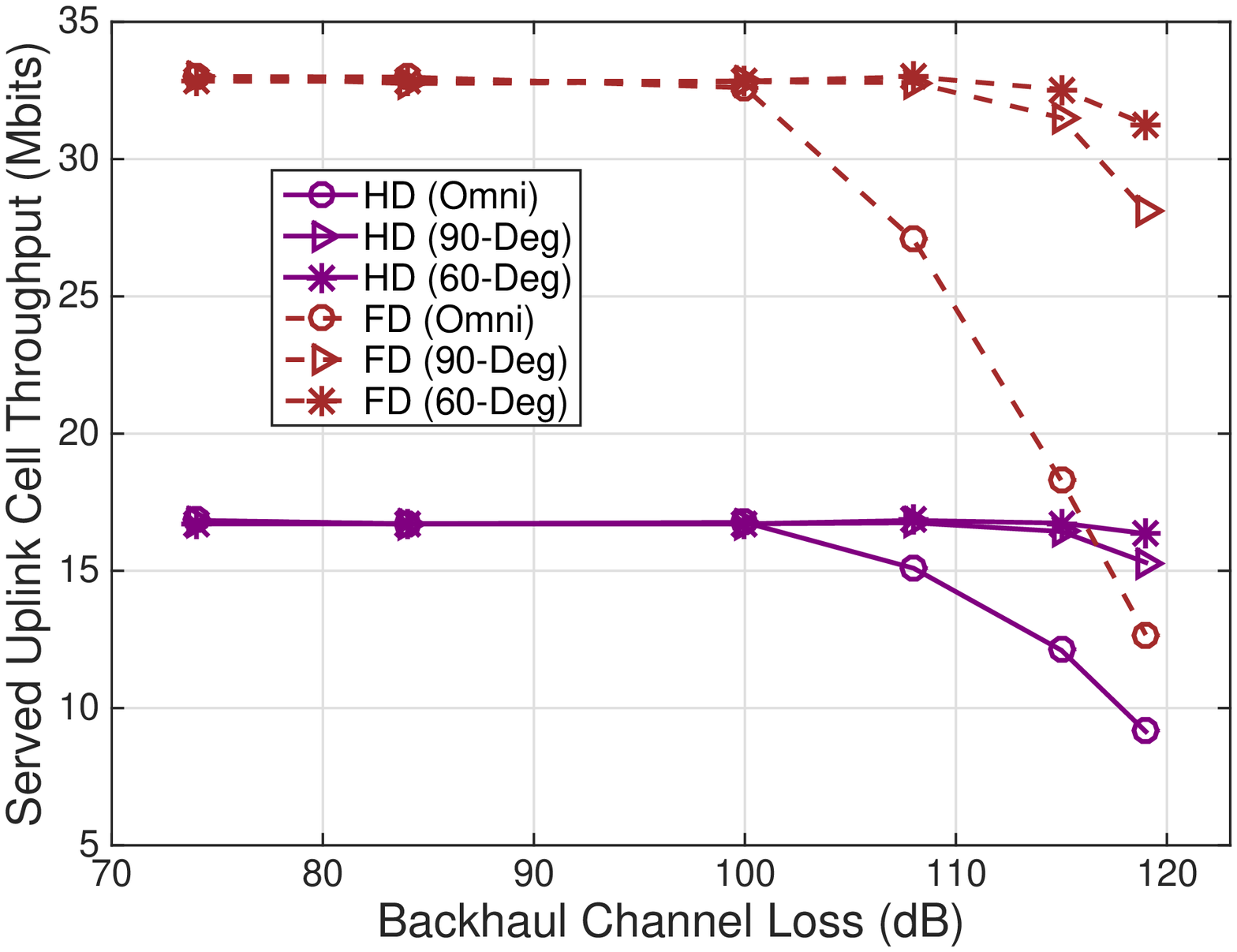}
	\vspace{-4mm}
	\caption{Uplink}
	\label{fig:SB_up_servTh_dir_eqDeman}
    \end{subfigure}
    \vspace{-2mm}
    \caption{Served cell throughput with directional transmissions.}
     \vspace{-5mm}
\end{figure}

The results above show that FD gain is small when the small cells are far from the macro BS, due to the weaker backhaul links. To improve the backhaul link quality, we also studied the effect of directional transmissions on the backhaul links. We assume that macro BS and the small cell BS have directional antennas, pointing at each other, for backhaul traffic. At the same time, they can transmit simultaneously to UEs using an omni-directional antenna. All UEs continue to have omni-directional antennas. We generated results for both 90\degree and 60\degree directional antennas. Figures~\ref{fig:SB_down_servTh_dir_eqDeman} and~\ref{fig:SB_up_servTh_dir_eqDeman} compares the results with different cases. It is clear that directional transmission strengthens the backhaul channels, which in turn improves the overall cell throughput. With directional antennas, the proposed solution is capable of doubling cell throughput even if the small cell is far from the center of the network, at the edge of the macrocell.

\vspace{-2mm}
\section{Conclusion}\label{sec:SB_conc}
In this work, we extend the application of FD radios to self-backhauled small cells, where the FD operations enables simultaneous backhaul and access uplink/downlink transmissions, as well as simultaneous uplink and downlink access/backhaul transmissions. We did a capacity analysis to compare the performance of the system with FD and HD operations under different propagation conditions. Using FD radios at both the small cell BS and macro-cell BS, we proposed a interference-aware hybrid scheduler, which jointly schedules the appropriate links  and allocates powers to maximize the gain of all the UEs in both uplink and downlink directions. Based on the traffic demand to it needs to satisfy and the interference condition, it switches between FD and HD modes. Our simulation results show that a full duplex radio can improve the capacity compared to half duplex systems by nearly two times in both directions. As an extension of this work, we are considering the performance of full duplex radios in a multi macro-cell scenario, each cell with multiple small cells.
\vspace{-3mm}
\bibliographystyle{IEEEtran}
\bibliography{FD_references}

\end{document}